# Unified Growth Theory: A puzzling collection of myths based on hyperbolic illusions

**Ron W Nielsen**[1]

Environmental Futures Centre, Gold Coast Campus, Griffith University, Qld, 4222, Australia

December, 2013

The Unified Growth Theory is a puzzling collection of myths based on illusions created by hyperbolic distributions. Some of these myths are discussed. The examination of data shows that the three stages of growth (Malthusian Regime, Post-Malthusian Regime and Modern Growth Regime) did not exist and that Industrial Revolution had no influence on the economic growth and on the growth of human population. All elaborate explanations revolving around phantom features created by hyperbolic illusions might be fascinating but they are scientifically unacceptable and, consequently, they do not explain the economic growth. The data clearly indicate that the economic growth was not as complicated as described by the Unified Growth Theory but elegantly simple.

> "If the proof starts from axioms, distinguishes several cases, and takes thirteen lines in the text book … it may give the youngsters the impression that mathematics consists in proving the most obvious things in the least obvious way"— George Pólya, Hungarian mathematician.

**Introduction**

Oded Galor, the world renowned economist, spent twenty years (Dinopoulos, 2012) developing his Unified Growth Theory (Galor 2005, 2011). However, his theory is a puzzling collection of phantom features based on hyperbolic illusions. Puzzling because it is so easy to reveal them as phantom features and puzzling because it is hard to understand why Galor did not see them for what they were. Puzzling because it is hard to understand why Galor was repeatedly presenting crudely constructed figures magnifying the illusions (see Fig. 1). Puzzling because it is hard to explain why he did not follow the standard practice of data investigation by using different ways of displaying them. The data do not change when they are displayed in different ways but certain features, which are otherwise hidden or obscure, can be revealed.

---

[1] (AKA Jan Nurzynski) r.nielsen@griffith.edu.au; ronwnielsen@gmail.com;
http://home.iprimus.com.au/nielsens/ronnielsen.html

Could Galor have saved twenty years of his life by spending less than an hour in carrying out a simple analysis of data? Could he have saved many years of life of those who study his theory and who believe that they can learn something useful?

When we examine closely his theory, we can identify at least the following myths based on hyperbolic illusions:

1. The myth of Malthusian stagnation.
2. The myth of the three regimes of growth, the concept developed from the myth of Malthusian stagnation.
3. The myth of lift-offs, particularly the lift-offs in the per-capita economic growth.
4. The myth of differential lift-offs, i.e. lift-offs occurring at different time for different regions. The myth of "differences in the timing of the take-off from stagnation to growth" (Galor, 2005, p. 178)
5. The myth that the Industrial Revolution influenced the economic growth.
6. The myth of the so-called great divergence

It is amazing how much time and space has been devoted to the explanation of the mechanisms of these phantom features, the process showing that even non-existing attributes can be flavoured with academic aroma and offered as a consumable product. We shall now examine some of these myths using the data for the world economic growth.

**The data and the fundamental questions**

Economic growth is measured by using the Gross Domestic Product (GDP) or equivalently by the GDP per capita (GDP/cap). Galor's fundamental idea is that the economic growth and the growth of human population were in three stages dividing the growth into three regimes: Malthusian Regime, Post-Malthusian Regime and Modern Growth Regime. This idea was motivated by observing that the GDP and the size of the population were small for hundreds of years but increasing rapidly in the recent time.

The contrast between the slow and fast growth is particularly dramatic if we plot the GDP/cap data because for this ratio the contrast is stronger than for the GDP data, the effect that can be easily explained using the properties of mathematical distributions. Furthermore, this contrast can be made even more dramatic by displaying the data in a certain way. The simple prescription is to select just a few strategically located points, preferably only four, form the larger set of data, to join these selected points by straight lines and to display the constructed figure using a linear set of coordinates, the method employed repeatedly by Galor. A sample of his display is shown in Fig. 1. He also used the same method for displaying regional data.

In the form displayed by Galor, the data give a compelling support for his theory. Fig. 1 shows clearly a long era of stagnation followed by a clear lift-off to a rapid economic growth. However, such an interpretation becomes questionable if we include other data points in our display and if we change the scales from linear to semilogarithmic (see Fig. 2). This new display seems to confirm Galor's observations but now it also raises certain questions.

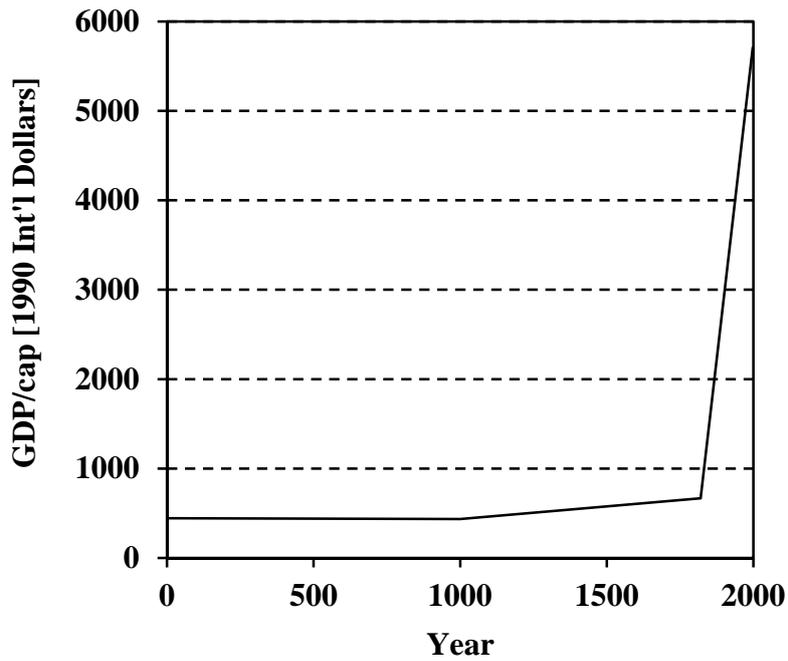

Fig. 1. The impoverished display of Maddison's data (Maddison, 2001) describing the world economic growth. The figure was reproduced from Galor's description of his Unified Growth Theory (Galor, 2005, p. 181). In this form, the data present a compelling case for the postulate of a prolonged stagnation followed by a rapid lift-off to a new regime of growth.

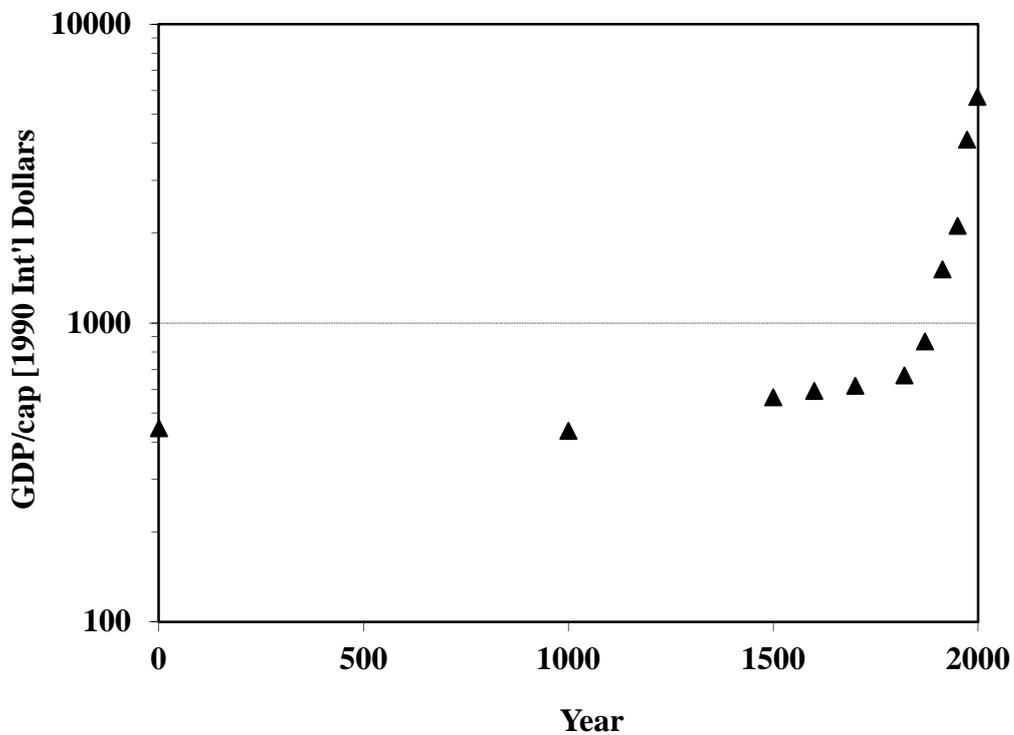

Fig. 2. The display of all Maddison's data (Maddison, 2001) describing the world economic growth. The rapid transition from slow to fast growth is now uncertain.

There is absolutely no question that the GDP/cap values were small for a long time. There is also no doubt that they were increasing rapidly in recent years. However, the important question is how rapid was this apparent transition from the slow to fast growth. Indeed, the question is whether there was a transition. With the exception of one or two points, the data seem to suggest that there was a seamless continuation of the growth trajectory, perhaps without any transition. This feature, therefore, deserves further investigation. It is an important investigation because it might decide between life and death of the theory.

According to Galor, the Malthusian Regime lasted for hundreds of years and was characterised by a stagnant stage of growth. On the other hand, the modern growth was explosively fast. Between these two extreme regimes there was the Post-Malthusian Regime. This description applies both to the growth of human population and to the economic growth.

If we accept these concepts, then the intriguing question, which was confronting Galor, is how to explain these three regimes. Why was the economic growth so slow for such a long time and why is it so fast now? Each of these three regimes must have been controlled by different sets of forces. What forces were controlling the economic and demographic growth in the past and what forces were controlling the growth in recent years? What triggered the sudden and dramatic escape from the tyranny of the apparently strong trap constraining the economic and demographic growth in the past and when did this escape happen?

If we are concentrating our attention on the features displayed in Fig. 1, we are missing a very important point. The fundamental question is not whether the growth was fast and slow but whether it was in three stages, governed by different sets of forces in each stage, with additional forces causing transitions between these three stages, such as the force of the Industrial Revolution. Is it possible that what we see as the growth in two or three stages is in fact the growth in just one, continuous and uninterrupted stage?

If the economic growth was in three stages, Galor was right and his theory stands firm. However, if the growth was in one stage, he was wrong and his Unified Growth Theory does not describe the economic growth.

If the three stages of growth did exist, all the questions we have asked earlier, and all other relevant questions we might add to them, make sense and it is then both interesting and important to read and study how Galor managed to answer them. However, if there was just one stage of growth, all these questions, and any other relevant questions we might have added to them are completely irrelevant. All the answers and explanations are also irrelevant and we can save a lot of time and effort if we ignore the Unified Growth Theory, except perhaps if we want to try to understand how a complicated theory can be created from nothing.

If there was just one stage, then the only pertinent and important question is why there was just one stage of growth. This question leads then away from the Unified Growth Theory to a truly unified growth theory because if there was just one stage of growth then there is no need to unify this one stage but only to explain why it was so unified.

So, now, how can we check whether the economic growth was in three stages or in just one stage? How could have Galor saved his time? How could he have saved the time of those who study his theory? The answer is remarkably simple: use the reciprocal values of data. This clue requires an explanation.

**Analysis of the data**

Anyone familiar with hyperbolic distributions will see immediately that the data of Maddison (2001) shown in Fig. 2, or even displayed in its crude form in Fig. 1, appear to follow a hyperbolic trajectory. However, those who are less familiar with mathematical distributions will just as easily see two components: slow and fast, stagnant and explosive. Consequently, such distributions have to be treated with care because they can be strongly misleading. Without a careful analysis they can easily lead to the formulation of various theories based on nothing else but illusions. Taking just four points from such distributions, plotting them using linear coordinates and joining the points by straight lines, as Galor did, is a perfect prescription for drawing incorrect conclusions.

One of the ways to try to check whether a hyperbolic-like distribution is made of two or three components is to plot data using semilogarithmic set of scales as we have done in Fig. 2, but an even better way is to plot the reciprocal values of data because if the data follow the increasing first-order hyperbolic distribution, their reciprocal values will follow a decreasing straight line. In our case, the reciprocal values of the size of human population and the reciprocal values of the GDP should follow decreasing straight lines. If data increase along a higher order hyperbolic trajectory, the reciprocal values will follow a monotonically decreasing but non-linear trajectory.

In order to understand the GDP/cap data we have to investigate separately the growth of the GDP and the growth of the population. The reason is that if the GDP and the size of the population follow hyperbolic distributions, the GDP/cap represents a ratio of hyperbolic distributions which on its own is hard to unravel.

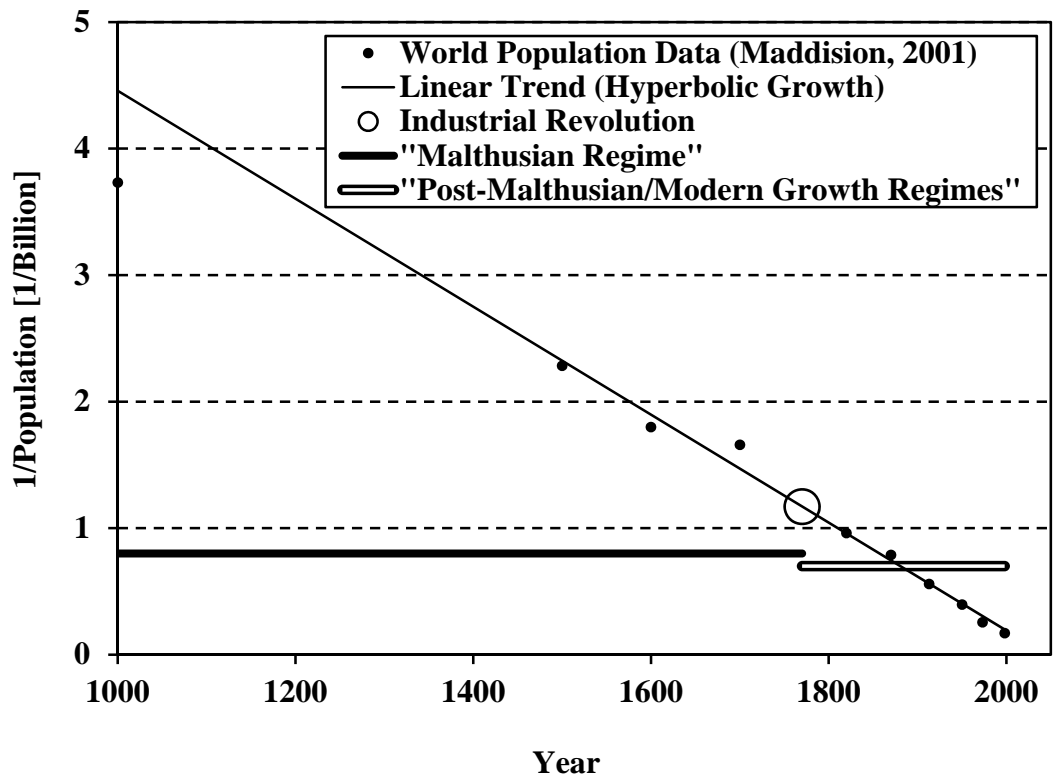

Fig. 3. Reciprocal values of the size of the world population (Maddision, 2001) based on the date used by Galor (2005, 2011) to develop his Unified Growth Theory. They show that the three regimes of growth did not exist and that the Industrial Revolution made no impression on the growth of human population.

Fig. 3 displays the reciprocal values of the size of the world population. This figure shows that during the so-called stagnant Malthusian Regime, the reciprocal values of the size of human population were monotonically decreasing, which means that the size of human population was monotonically increasing. There is no sign of any irregular or chaotic growth and no sign of stagnation. *The stagnant Malthusian Regime did not exist.*

Fig. 3 also shows clearly and convincingly that there was no change in the growth trajectory, no transition from the so-called Malthusian Regime to the imaginary Post-Malthusian Regime, and no transition from the imaginary Post-Malthusian Regime to the imaginary Modern Growth Regime. The Industrial Revolution, repeatedly lauded as having a crucial influence on changing the pattern of growth of human population, made no change in the pattern of growth. It is as if for the growth of human population the Industrial Revolution did not exist. There was Industrial Revolution but it had no effect on the growth of human population. *The data show that the mythical three stages of growth (the Malthusian Regime, the Post-Malthusian Regime and the Modern Growth Regime) did not exist. The growth of human population was in one continuous stage.*

Now let us look at the reciprocal values of the GDP data (Fig. 4). They also clearly demonstrate that *the stagnant Malthusian Regime, the Post-Malthusian Regime and the Modern Growth Regimes did not exist and that the Industrial Revolution had no effect on the world economic growth.*

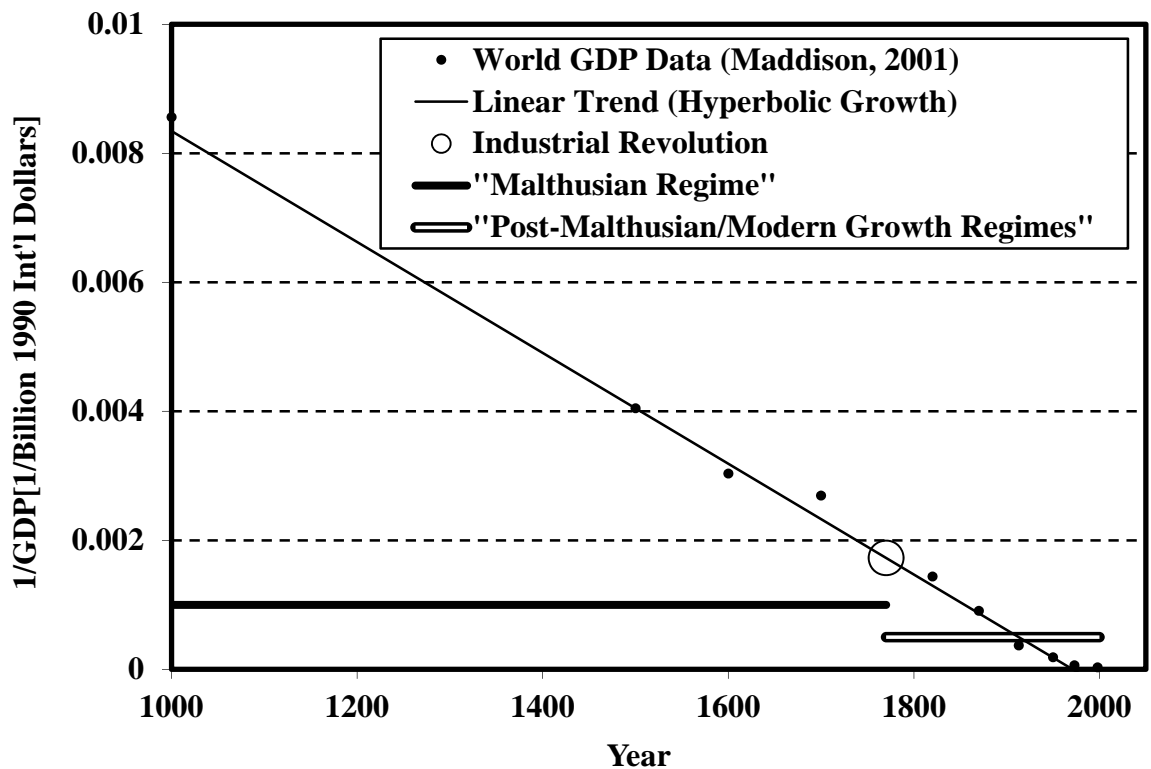

Fig. 4. Reciprocal values of the GDP data (Maddison, 2001), exactly the same as used by Galor (2005, 2011) to create his Unified Growth Theory. The data show that the three phantom regimes of growth, the three fundamental features described and "explained" by the Unified Growth Theory, did not exist. The Industrial Revolution did *not* have even the slightest effect on the world economic growth.

Figs 3 and 4 show that the size of human population and the GDP were increasing hyperbolically, because their reciprocal values can be described well by the decreasing straight lines. The last point for the reciprocal values of the GDP data is not aligned with the straight line because from the mid-1970s the world economic growth was slowing down.

Just by plotting the reciprocal values of data, the exercise requiring less than an hour, the Author of the Unified Growth Theory would have discovered that he was guided by illusions. This exercise is so simple that it might be considered too trivial. We might prefer to see complicated calculations or convoluted discussions to be convinced that the widely accepted theory is incorrect. However, trivial or not, the proof is undeniably strong and convincing: the three regimes of growth did not exist; the Industrial Revolution had no effect on the economic growth. It is only surprising that no-one else saw it earlier.

The GDP/cap is the ratio of the GDP and of the size of human population $S(t)$:

$$\frac{GDP}{cap} = \frac{GDP}{S(t)}$$

Considering that the size of human population, $S(t)$, and the GDP did not increase in three stages but in one, it is obvious that *the economic growth described by the GDP/cap data was not in three stages but in one. The three regimes of growth, the Malthusian Regime, the Post-Malthusian Regime and the Modern Growth Regime, did not exist.* There was no escape from the Malthusian trap because there was no trap. *Whatever the Unified Growth Theory is describing it is not describing the economic growth.* So we are back to where we started: the economic growth remains unexplained but at least we can now take a new direction and try to explaining it. If demographic and economic research is supposed to be treated as science, such incorrect claims should be readily and quickly corrected.

Our simple analysis answers the important and intriguing question: *Why was the economic growth so slow over hundreds of years and why was it so fast in the recent time?* These features are real. They are shown by the data. We cannot reject them just because we have to reject the Unified Growth Theory. The simple answer to this question is: *The economic growth was slow over hundreds of years and fast in the recent time because it was hyperbolic.*

We have shown in Figs 3 and 4 that both the growth of human population and the economic growth were hyperbolic because the reciprocal values of the size of human population and of the GDP follow the decreasing straight lines. We can make this answer perhaps even more explicit by displaying the GDP and GDP/cap data and their corresponding calculated trajectories (see Fig. 5).

The hyperbolic fit to the GDP data was calculated directly from the fit to their corresponding reciprocal values shown in Fig. 3. The fit to the GDP/cap data was calculated by dividing the hyperbolic fits to the GDP data and to the size of the population shown if Figs 3 and 4. The calculated curves follow closely the GDP and the GDP/cap data. The last point for the GDP and GDP/cap is outside the calculated hyperbolic trajectories because, as already mentioned, from the mid-1970s the world economic growth started to slow down.

The important point here is that the growth was not in two or three stages but in one over hundreds of years. The growth was not controlled by many sets of driving forces, different for each of the two or three imaginary regimes, but by just one force or by one set of forces. The continuing hyperbolic growth suggests that it was not as complicated and untidy as incorrectly described by the Unified Growth Theory. The important point also is that Golor's Unified Growth Theory did not produce any fits to the data.

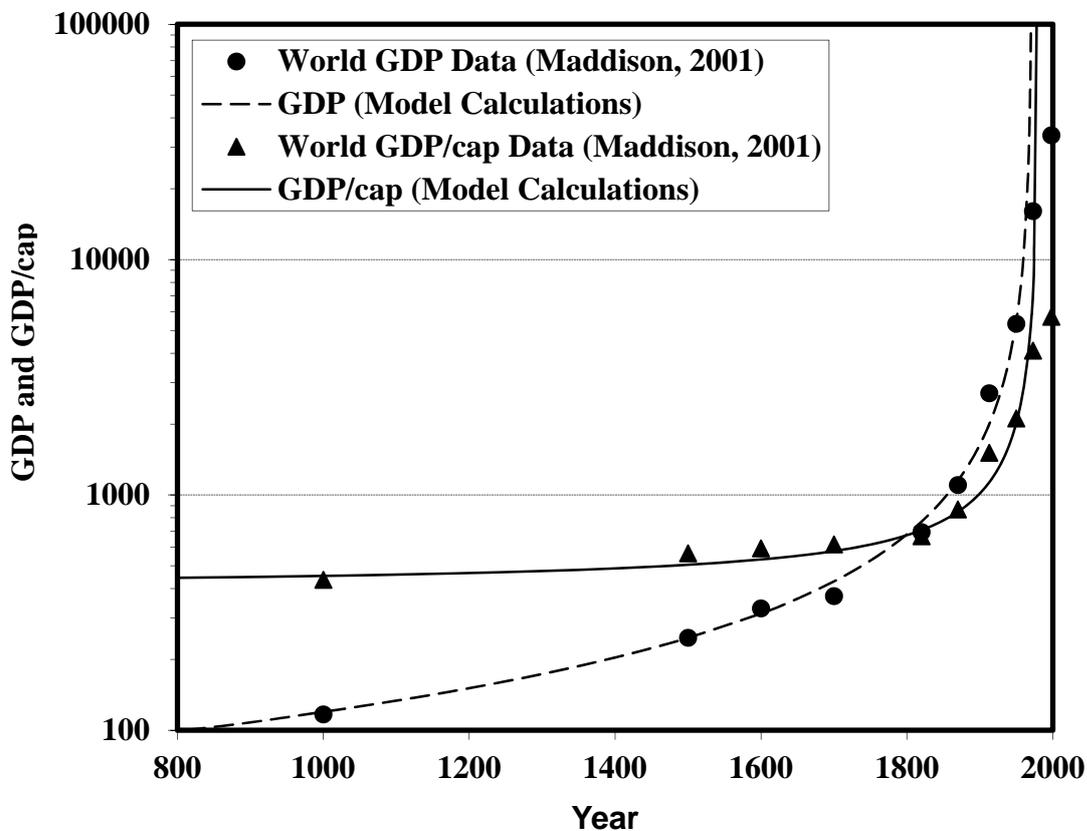

Fig. 5. The data of Maddison (2001) are compared with hyperbolic distributions calculated using the fits displayed in Figs 3 and 4. The three regimes of growth claimed by Galor (2005, 2011) did not exist. The GDP/cap data are in the 1990 International Dollars and the GDP in billions of the 1990 International Dollars.

So now, the pertinent question is not *Why was the growth in three stages?* – the question apparently asked by Galor, because he tries to answer it it in his theory, the irrelevant question because the three stages did not exist. The relevant question is *Why was the growth hyperbolic?* The related question is: *Why was the economic growth and the growth of human population so stable and so robust over such a long time?*

We can get further we ask correct questions. It also better to try first to understand data and only then to try to explain them rather than trying to explain the data without understanding them, as it was done in the Unified Growth Theory.

The last two questions and more are answered in the forthcoming book (Nielsen, n.d.). In this book we analyse not only the world economic growth but also the regional economic growth using the latest data of Maddison (2010). We explain why the hyperbolic illusion is exceptionally strong for the GDP/cap ratio. We also explain the illusion of the "great divergence," another incorrect concept belonging to the Unified Growth Theory, the illusion created partly by the properties of the hyperbolic growth and partly by the repeatedly impoverished display of data (Galor, 2005, 2008, 2011; Snowdon & Galor, 2008). In this book we also present an extensive discussion of human population dynamics.

## Conclusions

The display of the reciprocal values of data demonstrates convincingly that the size of human population and the GDP were monotonically increasing. Consequently, the GDP/cap was also monotonically increasing. The three regimes of growth, the stagnant Malthusian Regime, the Post-Malthusian Regime and the Modern Growth Regime, claimed by Galor (2005, 2011), did not exist. The fundamental assumptions of the Unified Growth Theory are contradicted by the relevant data well known to its creator. Whatever the Unified Growth Theory is describing or explaining it is not describing or explaining the economic growth.

Mathematical analysis of the data (Maddison, 2001), never attempted by Galor, clearly shows that the GDP data follow hyperbolic distribution, while the GDP/cap data follow the distribution representing the ratio of hyperbolic distributions. There is nothing mysteriously complicated about them and nothing so puzzling as to justify the creation of a complex theory, a misleading and irrelevant theory because it is contradicted by a rigorous analysis of data and because it tries to explain irrelevant phantom features creates by hyperbolic illusions.

Economic growth was slow in the past and fast in recent years because it was hyperbolic. There was no stagnation and no explosion but a steady hyperbolic growth. The continuous hyperbolic growth suggests a simple explanation of the mechanism of growth. Indeed, close analysis of data described in the forthcoming book (Nielsen, n.d.) shows that the economic growth was prompted by a *single* force of growth. The real world is not as untidy and as complicated as described by this Unified Growth Theory but elegant in its simplicity.